# Astrocytes: Arnol'd Tongues Generalization in Dynamical Systems' Parameter Plane

Gonzalo Marcelo Ramírez-Ávila,[1,2,3] S. Leo Kingston,[4] Marek Balcerzak,[2] Jérôme Daquin,[5] Timoteo Carletti,[1] and Tomasz Kapitaniak[2]

[1]*Namur Institute for Complex Systems (naXys),*

*Université de Namur, Rue de Bruxelles 61, B-5000 Namur, Belgium*

[2]*Division of Dynamics, Lodz University of Technology, 90-924 Lodz, Poland*

[3]*Instituto de Investigaciones Físicas, Universidad Mayor de San Andrés, La Paz, Bolivia*

[4]*Centre for Nonlinear and Complex Networks, SRM Institute of Science and Technology, Ramapuram, Chennai 600089, Tamil Nadu, India*

[5]*Univ Toulon, Aix Marseille Univ, CNRS, CPT, Toulon, France*

# Abstract

We discovered generalized structures, named astrocytes due to their shape, that constitute a defined region characterizing regular behavior within the parameter plane (PP) of dynamical systems (DSs). Morphologically, they are characterized by a branch and a soma with several vertices (arms) and sometimes with multiple periodicities. A bunch of infinite astrocytes emerge through their branches from a region, in general, of low periodicity. Astrocytes are embedded in a quasiperiodic-chaotic scenario. The soma complexity (number of vertices) determines a kind of hierarchy of the astrocytes; moreover, bunches of subsequent structures from the astrocyte have been emphasized, revealing a self-similarity property. We conducted a detailed analysis in a Zeeman laser model, but we also observed astrocytes in many other DSs. The multiperiodicity exhibited by the astrocytes in their soma gives rise to harlequin dress-like patterns and tri-, quad-, and quint-critical points, which indicate the coexistence of different higher-order periodicities. In the concave borders of the soma, a doubling cascade of quint-points emerges as a bifurcation in the PP, defining regions of ordered sequences of higher periodicity in the route to chaos.

Since the first efforts to disentangle the three-body problem initiated by Poincaré [1], the research field of dynamical systems (DSs) has evolved thanks to the works of numerous scientists, such as Lyapunov, van der Pol, Birkhoff, von Neumann, Kolmogorov, Lorenz, Shilnikov, Smale, Thom, Field, May, H´enon, Li, Yorke, Grebogi and Rössler, to mention a few; one of the main conclusions arising from those works is that chaotic behavior is recognized as common and ubiquitous in diverse systems [2–4]. On the other hand, concepts such as self-sustained oscillator [5] and attractor [6] played an important role in the theory of nonlinear oscillations in which regular oscillatory behavior comprises periodic oscillatory behavior and quasiperiodicity, namely dynamical states characterized by oscillations containing at least two incommensurable frequencies. More specifically, quasiperiodicity can be seen as a possible route to chaos and also as a dynamic state with competing incommensurate frequencies, whose ratio, called the winding number, is irrational [7, 8]. Let us observe that quasiperiodicity can also be found in a spatial context where it refers to the boundary separating periodic and aperiodic spatial distributions of the constituents of a material [9, 10]. The above-mentioned aspects have attracted the attention of scholars in the last four decades, the quasiperiodic route to chaos has been described in systems of diverse nature such as in convective fluids [11] and superfluids [12], in plasma [13], and even in transmission control protocols related to the Internet [14]. On the other hand, the characterization of quasiperiodicity has been carried out in different kinds of systems, such as in the peroxidase-oxidase reaction [15, 16], the Zeeman laser [17], an earthquake model [18], a population model [19], and a discrete neuron model [20, 21] among others.

The first representations of the dynamical behavior in the parameter plane (PP) are due to Arnol'd [22], where it is possible to distinguish regimes of periodic oscillations from quasiperiodicity by means of the stability regions, the so-called Arnol'd tongues, which are related to "horn-like" or "kite-like" shapes. Subsequently, the analysis in the PP conducted to the discovery of several stability regions immersed into chaotic or quasiperiodic regions such as the "shrimps" [23, 24] and their variants: spirals [25–27], networks [28–31], "eye(s)" of chaos [32], tricorns [33], rings [34], and "fishbones" [35] among others.

Among the tools used to characterize DSs, the analysis of PPs by using the quantification of the isospikes, i.e., number - per cycle - of maxima in mixed-mode oscillations (alternation of small and large amplitudes in a repeating pattern) [36] constitutes one of the most visually striking representations due to the distinguishability of the different kinds of regular oscillations on the PP of systems of diverse nature, ranging from physical [37–39] to chemical [40–42] and biological [43, 44], which were studied under this approach. Recently, scholars have brought to the fore an interesting dynamical feature present in the PP of several systems, namely, the existence of quint critical points. At each of these points, five distinct regular dynamic states coexist, each characterized by a different number of isospikes per cycle. This phenomenon was reported in electronic circuits [45, 46], in forced chemical [47, 48] and mechanical [49] systems, in free-running [50] and optically injected vertical-cavity surface-emitting lasers (VCSEL) [31], in a damped-driven curved carbon nanotube oscillator model [51] and in an intensity-modulated magneto-optical trap model [52]. In most cases, the authors found a form of chirality in the distribution of regions characterized by the number of isospikes and also noted the existence of a lattice of such quint points. Note that, in general, the systems studied are forced or quite complex, as in the case of lasers.

After a flourishing development of laser theory, some works dealing with lasers with an isotropic cavity, such as [53], gave rise to the so-called Zeeman laser model, which has been studied in different contexts of nonlinear dynamics, such as the on-off intermittency [17] and the quasiperiodic behavior [54]. Recently, some authors have revisited the Zeeman laser model, finding interesting results related to extreme events due to instabilities and quasiperiodicity [55, 56]; they also described essential aspects in the transition to hyperchaos [57]. The Zeeman laser might be described as a laser model with a cavity medium containing lower and upper levels $J = 0 \leftrightarrow J = 1$, which interacts with a monochromatic electric field, considering a strong anisotropy in the cavity in such a way that other anisotropies might be negligible. Consequently, the dynamic variables describing this laser model are the linear polarization components of the electric field $E_x$ and $E_y$. The transition coherence $P_k$ and the atomic inversion $D_k$ (where $k$ denotes the components, $k = x,y$) are associated with the transition between the upper sublevel $|k\rangle \equiv |J = 1, J_k = 0\rangle$ and the lower level $J = 0$. Ultimately, there exists the variable $Q$ representing the coherence between the sublevels $|x\rangle$ and $|y\rangle$. Consequently, the Zeeman laser behavior can be described by the following ordinary differential equations:

$$\begin{aligned}
\dot{E}_x &= \sigma(P_x - E_x), \\
\dot{E}_y &= \sigma(P_y - \alpha E_y), \\
\dot{P}_x &= -P_x + E_x D_x + E_y Q, \\
\dot{P}_y &= -P_y + E_y D_y + E_x Q, \\
\dot{D}_x &= (r - D_x) - 2(2E_x P_x + E_y P_y), \\
\dot{D}_y &= (r - D_y) - 2(2E_y P_y + E_x P_x), \\
\dot{Q} &= -Q - (E_x P_y + E_y P_x).
\end{aligned} \quad (1)$$

Here, $\alpha$ relates to the cavity anisotropy, while $\sigma$ and $\alpha\sigma$ denote the cavity losses in the $x$ and $y$ directions, respectively; $r$ represents the incoherent pumping rate, which is assumed to be identical across the $|k\rangle$ sublevels.

In this Letter, we present a dynamical analysis of the Zeeman laser, focusing on a quasiperiodic region in the PP, that allowed us to determine and describe the presence of new dynamical structures supporting regular dynamics. Due to their shapes and similarity with those found in the central nervous system [59], we denominate the latter as *astrocyte*. The proposed analysis also permitted us to realize that astrocytes constitute a generalization of Arnol'd tongues, but with higher complexity. The present study unveils the basic aspects of these structures, which consist of a *branch* and a *soma*, with their complexity being proportional to the number of vertices and, consequently, to the number of concavities. We describe the morphology of astrocytes (branch, soma, vertices, and self-similar extensions) and their relationship to the dynamical environments (quasiperiodic, chaotic, and hyperchaotic) as well as the inner feature of quint point sequences. Our results were obtained by integrating Eqs. (1) by using the Runge-Kutta-4 method with fixed initial conditions $(E_{x0}, E_{y0}, P_{x0}, P_{y0}, D_{x0}, D_{y0}, Q_0) = (0.2, 0.4, 0.3, 0.6, 1.05, 0.05, 0.25)$, by considering an integration time $t = 20 \times t_{\text{trans}}$, a transient, $t_{\text{trans}} \approx 1.1 \times 10^6$ and a time step $\Delta t = 0.01$. Without loss of generality, we decided to analyze the physical variable $E_y$, and our findings are grounded on counting the number of isospikes, i.e., the number of maxima contained in a period; the latter are complemented with other techniques to detect quasiperiodicity or distinguish chaotic from regular behavior, namely, the computation of the Lyapunov spectrum [60, 61] using variational methods [62] and other indicators such as the Mean Ex-ponential Growth Factor of Nearby Orbits (MEGNO) [63, 64] and their precursors [65] (see the evolution of some of these quantities for specific examples in Fig. S4 of [58]). Let us start by indicating that Kingston et al. [55] obtained a phase diagram in terms of parameters $r$ and $\alpha$ of the Zeeman laser model, allowing them to distinguish chaotic from quasiperiodic dynamics and even to identify quasiperiodic breakdown (QPB) and quasiperiodic intermittency (QPI) regions. In Fig. 1(a), we display the same phase diagram as shown in [55] (see Fig. S1 [58] obtained in terms of isospikes and the two largest Lyapunov exponents (LLEs)); we note some interesting details (see framed region that is zoomed in Figs. 1(a) and (b)). In principle one should expect them to contain mainly quasiperiodicity; surprisingly enough, we determined a bunch of structures with different shapes and exhibiting periodic behavior (crescent, astrocytes, and Arnol'd tongue-like structures). The crescent is characterized by the fact that it is almost completely submerged in a chaotic environment, and the other structures contain a branch emerging from the vast pink region describing sine-type oscillations (one peak per cycle), and every branch gives rise to a soma displaying several colored regions representing the number of isospikes. The presence of vertices (arms) is the other feature of these astrocytes. As shown in Fig. 1(b), the astrocyte structures exhibit a variety of shapes; the periodicity in their branches increases with the decrease in the values of the parameters $r$ and $\alpha$. Even though there are many structures indicating regular behavior in the bunch, the most visible are what we call astrocytes 1 (Fig. 1(c)) and 2 (Fig. 1(e)), characterized by branches of 13 and 17 isospikes per cycle, respectively. For parameter values inside the astrocytes, the system exhibits regular behavior, whose isospike number is reported with a distinctive color, creating a harlequin dress-like pattern in the soma of the majority of the structures. Let us observe that to a non-periodic dynamic, we assign a black point without distinguishing whether the dynamic behavior corresponds to quasiperiodicity or chaos. A complete dynamical description must permit distinguishing regular behavior from all the other possible dynamical behaviors: quasiperiodicity, chaos, and hyperchaos, if they exist. In order to complete the dynamic description of the inner and outer parts of the astrocytes, the Lyapunov spectrum is obtained, and the combination of the two LLEs allows us to represent astrocytes 1 and 2 and their surroundings in terms of such quantifiers (Figs. 1(d) and 1(f)), respectively. These diagrams reveal the regions exhibiting regular (black and white with $\lambda_1 \approx 0$, $\lambda_2 < 0$), quasiperiodic (olive green with $\lambda_1 \approx 0$, $\lambda_2 \approx 0$), chaotic (colored with $\lambda_1 > 0$, $\lambda_2 < 0$), and hyperchaotic (pink with $\lambda_1 > 0$, $\lambda_2 > 0$) behavior (in what follows, the color bar quantifies the LLEs). Moreover, in these figures we marked clockwise the most prominent vertices and found 11 and 6 in astrocytes 1 and 2, respectively. The number of vertices is proportional to the number of concavities. Notably, all the vertices are situated in the chaotic region. Additionally, a type of degeneration appears to occur when a vertex is in the hyperchaotic region, as illustrated in Fig. 1(d). Furthermore, the Arnol'd tongue-like structures exhibit similar characteristics regarding the vertices located in the chaotic region, even though the number of vertices is fewer than those associated with astrocytes. Periodicities (number of isospikes per period) and LEs approaches have their pros and cons, but an analysis combining both permits having a complete dynamical description of the astrocytes and their neighborhoods. In our analysis, we will focus on astrocyte 1. Additional information regarding this astrocyte, including the foundational aspects of the research, representative time series, return maps and attractors, the evolution of Lyapunov Exponents (LLEs) under various dynamical behaviors, and the emergence of self-similar structures related to astrocyte 1, is provided in Sec. S1. Furthermore, supplementary descriptions of the Arnol'd tongue-like structures, in addition to those for astrocyte 1, can be found in Sec. S2. The characteristics of astrocyte 2 and astrocytes within other dynamical systems are addressed in Secs. S3 and S4 of [58] respectively.

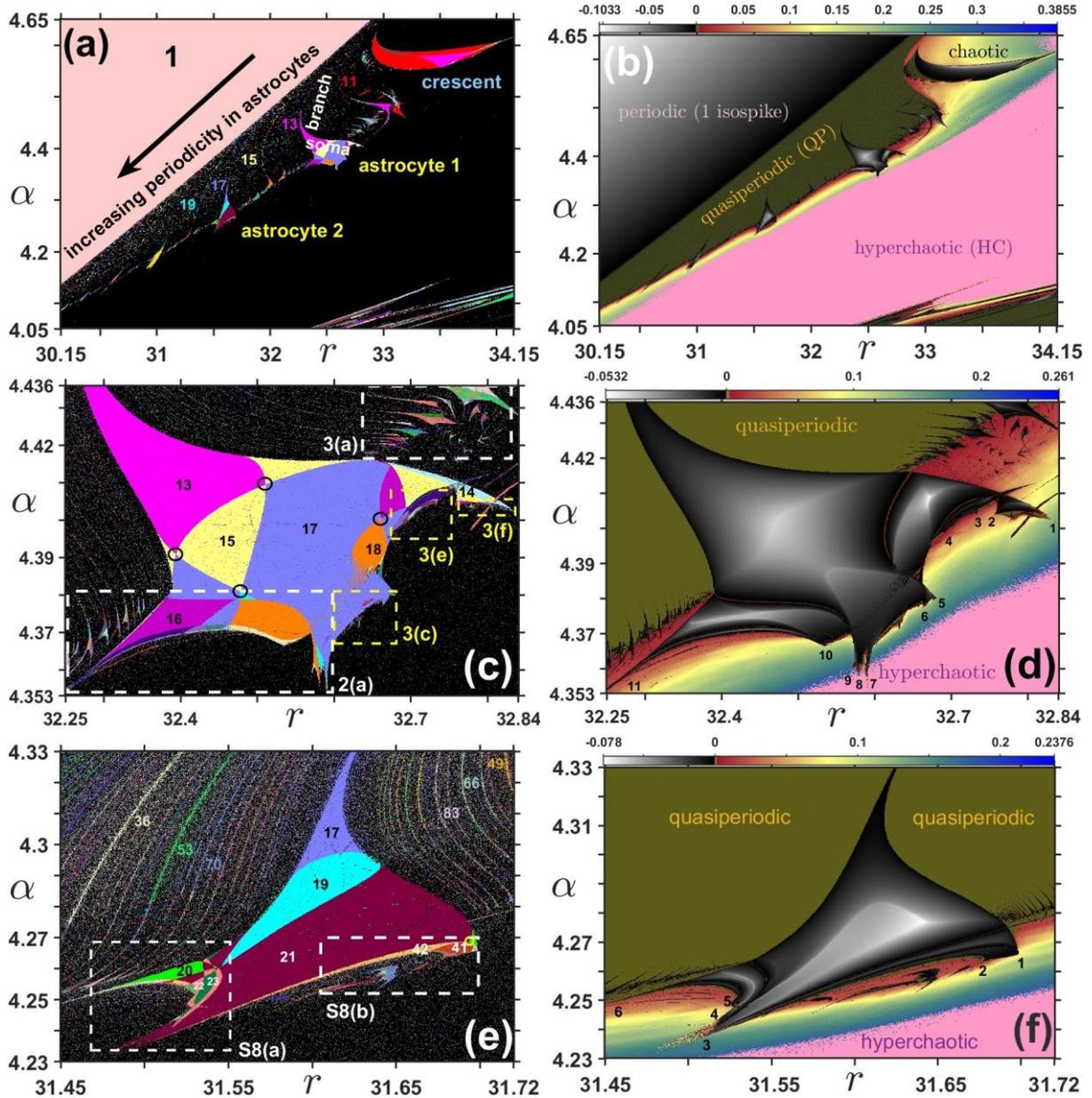

Figure 1: Parameter planes of the variable $E_y$ in terms of the number of isospikes (left column) and the two largest Lyapunov exponents (LLEs) (right column). (a) An enlarged view of the quasiperiodic region shown in [55] and in Fig. S1, where it is possible to distinguish a "crescent" structure and several astrocytes, particularly (c) 1 and (e) 2 with main periodicities given by 13 and 17 isospikes, respectively. Note the presence of critical points (tri, quart, or quint), indicated by circles. The framed regions in astrocytes 1 and 2 are studied in Figs. 2(a) and 3 and in Sec. S3 of the supplemental material [58], respectively. (b) (d), and (f) are representations in terms of the Lyapunov spectrum, where the two LLEs are combined to identify all the dynamical behavior exhibited inside and outside the astrocytes 1 and 2, respectively.

The upper part of Fig. 2 corresponds to a framed region of Fig. 1(c), illustrating the downward concavity of astrocyte 1 and its surroundings. This is analyzed by using the isospikes approach, along with the LEs presented in Figs. 2(a) and (b). According to Fig. 2(a), in the inner part of astrocyte 1, a quart point appears (black circle) where oscillations with 16 to 19 isospikes coexist. This gives rise to the first layer, $L_1$, which is composed of regions with 16, 17, and 18 isospikes per period. This leads to two quint points (yellow circles) and the second layer, $L_2$, which contains regions with isospikes ranging from 32 to 36. Fig. 2(c) provides more detail, exhibiting four quint points and the third layer, $L_3$, characterized by periodic regions containing isospikes from 64 to 72. Fig. 2(d) specifies this, also showing the eight quint points that precede the fourth layer, $L_4$, with a sequence of periodic regions containing 128 to 144 isospikes. Finally, Fig. 2(e) represents a part of the fifth layer, $L_5$, along with 12 of the 16 quint points. Thus, we observe that in the downward concavity region of astrocyte 1, there is a *number of isospikes doubling route to chaos*, but also a *quint-point doubling route to chaos*. The sequence in which the number of isospikes is present in the different layers is shown in the diagram of Fig. 2(f), where the range of the number of isospikes in the distinct regions of any layer $L = n$ is given by $[2^{n-1}(P - 1), 2^{n-1}(P), 2^{n-1}(P + 1)]$, while the number of quint points scales as $2^L$.

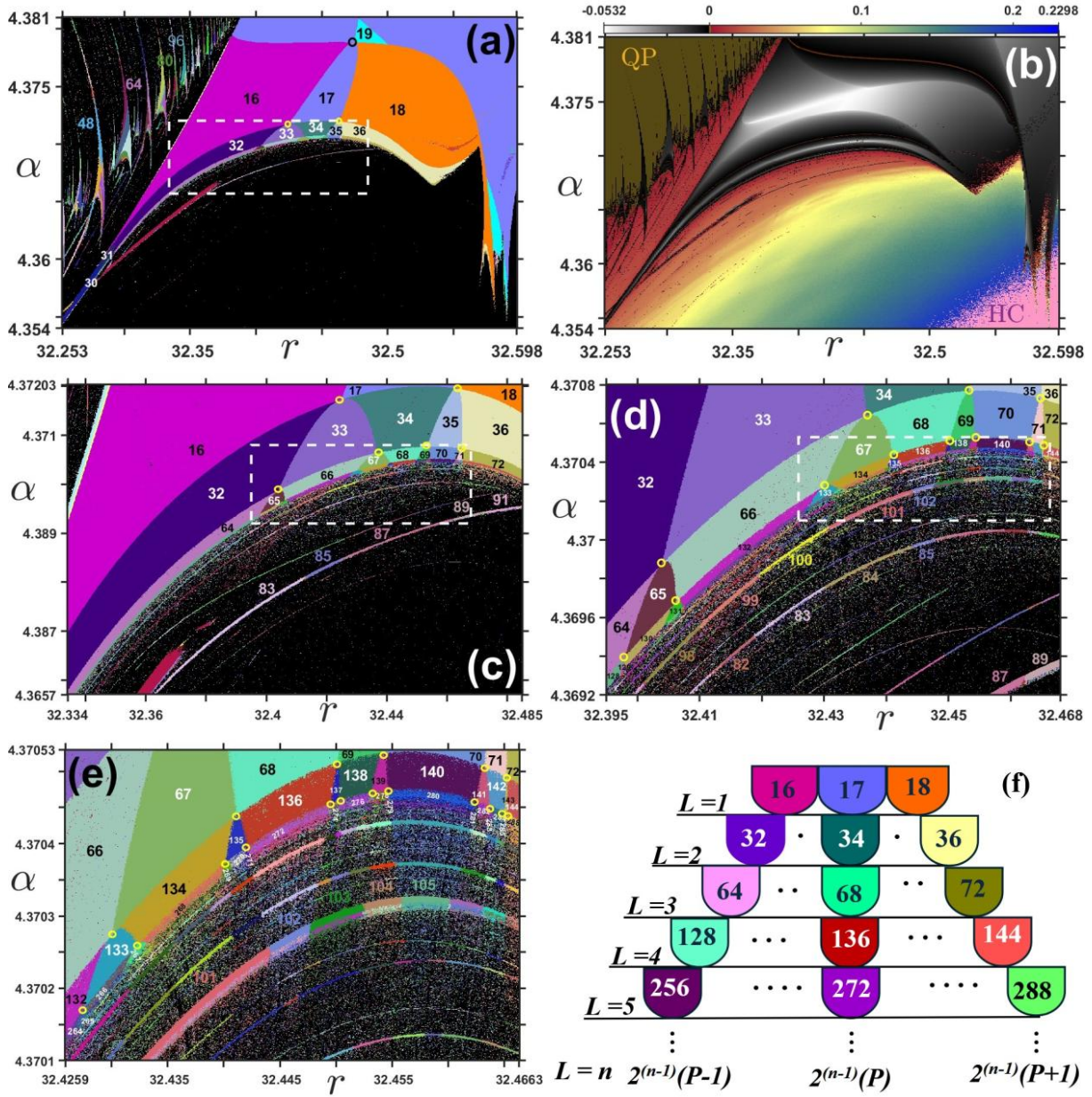

Figure 2: Enlarged view of the framed region shown in the lower-left part of Fig. 1(c); parameter planes in terms of (a) the number of isospikes and (b) the LLEs. Zoomed views of the framed areas corresponding to the successive layers and their quint points: (c) second (framed region of (a)), (d) third (framed region of (c)), (e) fourth (framed region of (d)). (f) Representation of the sequence in the number of isospikes per cycle in the successive layers.

On the other hand, in the left outer region of astrocyte 1, there is a sequence of Arnol'd-tongue-like structures, whose periodicity distribution increases towards the upper part in the more noticeable part of the branches in multiples of 16, and there are also the cascades of period-adding as those shown in [66, 67]; and in the outer lower part, there are sequences of periodic arcs with an increasing number of isospikes as far as $r$ increases. Whereas the representation with the LLEs of Fig. 2(b) allows us to confirm that all the vertices, both from the astrocytes and outer structures, are related to a chaotic region.

The upper-right framed region of Fig. 1(c) is represented in terms of the number of isospikes and LLEs in Figs. 3(a) and (b), respectively. We note that, besides the convex upward region of astrocyte 1, there is a sequence of kite-like structures that exhibit several periodicities within the soma. We also observe the presence of shrimps and other periodic structures embedded in the chaotic region. The framed region in Fig. 3(a) is represented in Fig. S7 of [58] where the sequence of quint points shows a case of chirality and its symmetry breaking. The emergence of self-similar structures emanating from different regions of astrocyte 1 are represented in Fig. 3(c)–(f). In Fig. 3(c), there is a bunch of Arnol'd tongues accompanied by a similar structure to the crescent mentioned in Fig. 1(a). The quasiperiodicity characterizing the bunch of Arnol'd tongues is shown in Fig. 3(d). Other bunches of Arnol'd tongues are shown in Fig. 3(e) emanating from periodic regions of the astrocyte 1 with 32 and 33 on the sides of a concavity region. In Fig. 3(f), the emerging self-similar structures appear from periodic regions of the astrocyte 1 with 14 to 16 isospikes; here also, a kind of crescent appears. The representation in terms of the LLEs is shown in Sec. S1.D, Fig. S6. Finally, some examples exhibiting the presence of astrocyte bunches in two cases: a modified Hénon map [68] and a VCSEL [31] are shown in [58].

To summarize, in this work we prove the existence of astrocytes as generalized structures in the PP of DS. They are characterized by their morphology (branch, soma, and multivertices) emanating from a structure indicating regular dynamic behavior. Consequently, the Arnol'd tongues in their horn- or kite-shaped forms constitute a special case of astrocytes. The astrocytes indicate geometrically one of the origins of quasiperiodicity because the surrounding area of the branch and the part of the soma not containing any vertex are immersed in a region of quasiperiodicity that extends to cover the entire bunch of emerging astrocytes. Naturally, the quasiperiodic region is pervaded by branches and parts of the soma of all the emerging astrocytes. When computing the winding number inside the astrocytes (in our case, the ratio between the number of isospikes from the generating region and the number of isospikes from the astrocyte), it yields a rational number; by contrast, the region outside the astrocytes may correspond to quasiperiodicity (irrational winding numbers) or to chaos, where the astrocyte's vertices lie within the PP. Given that the set of irrational numbers is uncountably infinite, it is clearly easier to detect quasiperiodicity than periodicity in the region defined by the bunch representing the emerging astrocytes. Another important aspect is self-similarity, which means that astrocytes can generate subsequent offspring of astrocytes that have simpler structures in terms of size and shape, shorter branch lengths, fewer vertices, and more basic inner dynamical behaviors (characterized by a smaller variety of periodicities in the soma). This implies that hierarchical structures exist based on the generation to which the astrocytes belong. Consequently, it is possible to classify astrocytes based on their (i) *morphology* (branch length and soma size); (ii) *topology* (number of vertices in the soma); (iii) *dynamics* (periodic features of the astrocytes and the regions from which they emerge); and (iv) *generational aspects* (quantity of offspring they can produce and their generational stage). Outside the astrocytes, there are structures such as "arcs," "shrimps," "manta rays," and others immersed in a chaotic region. An indicator of the chaotic environment is the presence of vertices and degenerated vertices, which might indicate a hyperchaotic scenario. The number of vertices is related to the number of concavities by $N_{concavities} = N_{verices} - 1$. The concavities constitute scenarios for the route to chaos, which can arise through either an isospikes-doubling and isospikes-adding cascade or a sequence of quint-point doubling. Moreover, concavities may also represent the regions where self-similar structures emerge, giving rise to the formation of a new generation of astrocytes. The emergence of astrocytes seems to be accompanied by other primary structures, such as the "crescent"-like structures observed beside or in between the bunch of astrocytes. The characterization of these accompanying structures might shed more light on astrocyte genesis. Finally, we note that the quint-point doubling cascade has the same features as a period-doubling bifurcation; consequently, it is expected to exhibit the same universality in relation to the Feigenbaum constants. The demonstration of such universality constitutes an appealing computational challenge.

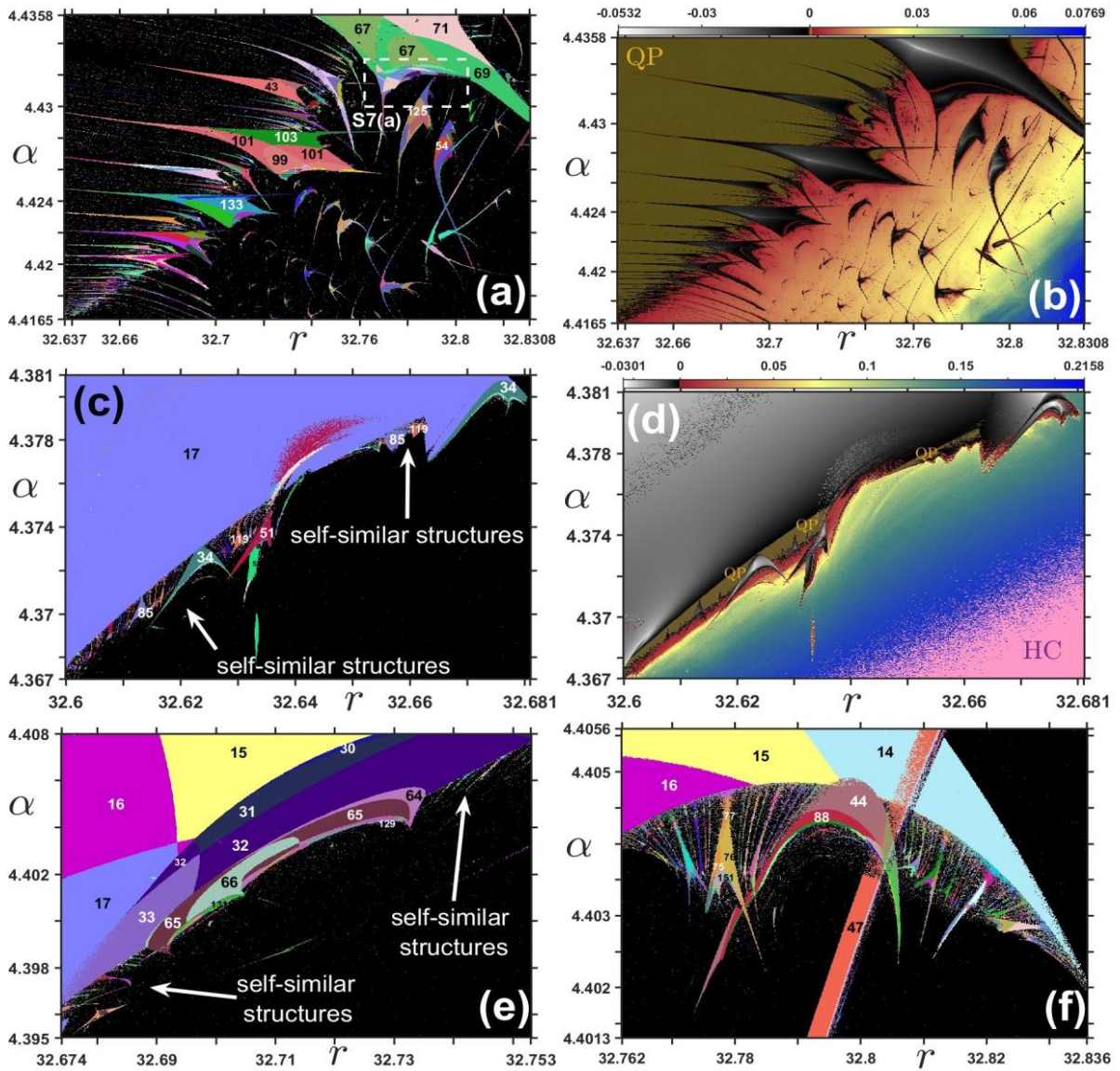

Figure 3: Enlarged view of the PPs framed regions of Fig. 1(c). Arnol'd tongues in the upper part of astrocyte 1 in terms of (a) the number of isospikes and (b) the LLEs, where it is noted the presence of structures indicating regular behavior lying in a chaotic region such as shrimps and other exotic shapes. Zoomed view of the other framed regions of Fig. 1(c) where in (c) and (d) it is possible to perceive the emergence of self-similar structures from a periodic region of the astrocyte 1 with 17 isospikes per period (c) and the quasiperiodic region (d). In (e), there are two other self-similar structures arising from the periodic region of the astrocyte with 32 and 33 isospikes per period. (f) shows two emerging self-similar structures from periodic regions of the astrocyte 1 with the number of isospikes 14, 15, and 16. The representation of PPs in terms of the LLEs of (e) and (f) is shown in Fig. S6 of the supplemental material [58].

This project has received funding from the European Union's Horizon 2020 research and innovation programme under the Marie Sklodowska-Curie grant agreement No 101034383. Computational resources have been provided by the Consortium des Equipements de Calcul Intensif (CECI), funded by the Fonds de la Recherche Scientifique de Belgique (F.R.S.-FNRS) under Grant No. 2.5020.11 and by the Walloon Region. T.K. acknowledges financial support by the National Science Centre, Poland, OPUS Program (Project No.2021/43/B/ST8/00641).

———

[1] H. Poincaré, Acta Math. **13**, 1 (1890).

[2] M. W. Parker, Stud. Hist. Philos. Sci. Part B: Stud. Hist. Philos. Mod. Phys. **29**, 575 (1998). [

[3] D. Aubin and A. Dahan Dalmedico, Hist. Math. **29**, 273 (2002).

[4] C. Letellier, R. Abraham, D. L. Shepelyansky, O. E. Rössler, P. Holmes, R. Lozi, L. Glass, A. Pikovsky, L. F. Olsen, I. Tsuda, C. Grebogi, U. Parlitz, R. Gilmore, L. M. Pecora, and T. L. Carroll, Chaos **31**, 053110 (2021).

[5] A. A. Andronov, C.R. Acad. Sc. Paris Math. **189**, 559 (1929).

[6] G. D. Birkhoff, *Dynamical systems*, Vol. 9 (American Mathematical Soc., Providence, 1927).

[7] R. Hilborn, *Chaos and nonlinear dynamics: an introduction for scientists and engineers* (Oxford University Press, Demand, 2000).

[8] J. C. Sprott, *Chaos and time-series analysis* (Oxford University Press, New York, 2003).

[9] M. Duneau and A. Katz, Phys. Rev. Lett. **54**, 2688 (1985).

[10] V. E. Korepin, J. Sov. Math. **41**, 956 (1988).

[11] J. Stavans, F. Heslot, and A. Libchaber, Phys. Rev. Lett. **55**, 596 (1985).

[12] R. Mainieri, T. S. Sullivan, and R. E. Ecke, Phys. Rev. Lett. **63**, 2357 (1989).

[13] D. Weixing, W. Huang, X. Wang, and C. X. Yu, Phys. Rev. Lett. **70**, 170 (1993).

[14] J.-B. Gao, N. S. V. Rao, J. Hu, and J. Ai, Phys. Rev. Lett. **94**, 198702 (2005).

[15] T. Hauck and F. W. Schneider, J. Phys. Chem. **97**, 391 (1993).

[16] L. Olsen, T. Bronnikova, and W. Schaffer, Phys. Chem. Chem. Phys. **4**, 1292 (2002).

[17] J. Redondo, E. Roldán, and G. J. de Valcárcel, Phys. Lett. A **210**, 301 (1996).

[18] O. Ramos, E. Altshuler, and K. J. Måløy, Phys. Rev. Lett. **96**, 098501 (2006).

[19] C. Rosa, M. J. Correia, and P. C. Rech, Chaos Solitons Fractals **40**, 2041 (2009).

[20] F. Wang and H. Cao, Commun. Nonlinear Sci. Numer. Simul. **56**, 481 (2018).

[21] G. M. Ramírez-Ávila, S. S. Muni, and T. Kapitaniak, Chaos **34**, 083134 (2024).

[22] V. Arnol'd, Izv. Akad. Nauk SSSR Ser. Mat. **25**, 21 (1961).

[23] J. A. C. Gallas, Phys. Rev. Lett. **70**, 2714 (1993).

[24] J. A. C. Gallas, Physica A **202**, 196 (1994).

[25] P. Gaspard, R. Kapral, and G. Nicolis, J. Stat. Phys. **35**, 697 (1984).

[26] C. Bonatto and J. A. C. Gallas, Phys. Rev. Lett. **101**, 054101 (2008).

[27] R. Barrio, F. Blesa, S. Serrano, and A. Shilnikov, Phys. Rev. E **84**, 035201 (2011), pRE.

[28] R. E. Francke, T. Pöschel, and J. A. C. Gallas, Physical Review E **87**, 042907 (2013).

[29] M. R. Gallas, M. R. Gallas, and J. A. C. Gallas, Eur. Phys. J.-Spec. Top. **223**, 2131 (2014).

[30] G. M. Ramírez-Ávila, I. M. Jánosi, and J. A. C. Gallas, Europhys. Lett. **126**, 20001 (2019).

[31] G. M. Ramírez-Ávila and T. Carletti, Chaos **35**, 033144 (2025).

[32] R. O. Medrano-T and R. Rocha, Int. J. Bifurcation Chaos **24**, 1430025 (2014).

[33] C. A. T. Chávez and S. Curilef, Chaos **30**, 023130 (2020).

[34] G. M. Ramírez-Ávila, J. Kurths, and J. A. C. Gallas, Chaos **31**, 101102 (2021).

[35] B. R. R. Boaretto, P. R. Protachevicz, M. Hansen, R. O. Medrano-T, E. E. N. Macau, and C. Grebogi Phys. Rev. E **111**, L052203 (2025).

[36] J. G. Freire and J. A. C. Gallas, Phys. Chem. Chem. Phys. **13**, 12191 (2011).

[37] L. Junges, T. Pöschel, and J. A. C. Gallas, Eur. Phys. J. D **67**, 149 (2013).

[38] J. A. C. Gallas, Eur. Phys. J. Spec. Top. **228**, 2081 (2019).

[39] S. Gu, P. Zhou, P. Mu, G. Guo, X. Liu, and N. Li, Opt. Express **31**, 31853 (2023).

[40] M. R. Gallas and J. A. C. Gallas, Chaos **25**, 064603 (2015).

[41] L. F. Olsen, Chaos **33**, 023102 (2023).

[42] G. M. Ramírez-Ávila, T. Kapitaniak, and D. Gonze, Chaos **34**, 073154 (2024).

[43] J. G. Freire, M. R. Gallas, and J. A. C. Gallas, Chaos **28**, 053118 (2018).

[44] G. M. Ramírez-Ávila, J. Kurths, D. Gonze, and G. Dupont, Chaos Solitons Fractals **173**, 113743 (2023).

[45] J. A. C. Gallas, Eur. Phys. J. Plus **136**, 1048 (2021).

[46] C. K. Volos and J. A. C. Gallas, Eur. Phys. J. Plus **137**, 154 (2022).

[47] J. A. C. Gallas, Phys. Chem. Chem. Phys. **23**, 25720 (2021).

[48] J. A. C. Gallas, J. Phys.-Condens. Mat. **34**, 144002 (2022).

[49] Z. Liu, X. Rao, J. Gao, and S. Ding, Chaos Solitons Fractals **177**, 114254 (2023).

[50] S. Gu, P. Zhou, and N. Li, Opt. Lett. **48**, 2845 (2023).

[51] C. Manchein, B. Fusinato, H. S. Chagas, and H. A. Albuquerque, Chaos **33**, 063147 (2023).

[52] A. L. de Oliveira, G. B. Corrêa, H. A. Albuquerque, K. M. Farias, and P. C. Rech, Phys. Lett. A **562**, 131023 (2025).

[53] G. P. Puccioni, M. V. Tratnik, J. E. Sipe, and G. L. Oppo, Opt. Lett. **12**, 242 (1987).

[54] J. Redondo, G. J. de Valcárcel, and E. Roldán, Phys. Rev. E **56**, 6589 (1997).

[55] S. L. Kingston, A. Mishra, M. Balcerzak, T. Kapitaniak, and S. K. Dana, Phys. Rev. E **104**, 034215 (2021).

[56] S. L. Kingston, S. Kumarasamy, M. Balcerzak, and T. Kapitaniak, Opt. Express **31**, 22817 (2023).

[57] S. L. Kingston, T. Kapitaniak, and S. K. Dana, Chaos **32**, 081106 (2022).

[58] See Supplemental Material .

[59] M. V. Sofroniew and H. V. Vinters, Acta Neuropatho. **119**, 7 (2010).

[60] A. Wolf, J. B. Swift, H. L. Swinney, and J. A. Vastano, Physica D **16**, 285 (1985).

[61] A. Pikovsky and A. Politi, *Lyapunov exponents: a tool to explore complex dynamics* (Cambridge University Press, 2016).

[62] C. Skokos, The Lyapunov characteristic exponents and their computation, in *Dynamics of Small Solar System Bodies and Exoplanets*, edited by J. J. Souchay and R. Dvorak (Springer Berlin Heidelberg, Berlin, Heidelberg, 2010) pp. 63–135.

[63] N. P. Maffione, C. M. Giordano, and P. M. Cincotta, Int. J. Nonlin. Mech. **46**, 23 (2011).

[64] P. M. Cincotta and C. M. Giordano, Theory and applications of the mean exponential growth factor of nearby orbits (megno) method, in *Chaos Detection and Predictability*, edited by C. Skokos, G. A. Gottwald, and J. Laskar (Springer Berlin Heidelberg, Berlin, Heidelberg, 2016) pp. 93–128.

[65] G. Benettin, L. Galgani, A. Giorgilli, and J.-M. Strelcyn, Meccanica **15**, 9 (1980).

[66] C. Bonatto and J. A. C. Gallas, Phys. Rev. E **75**, 055204 (2007).

[67] F. Augusto Cardoso Pereira, E. Colli, and J. Carlos Sartorelli, Chaos **22**, 013135 (2012).

[68] J. A. C. Gallas, Phys. Rev. E **48**, R4156 (1993).

# Supplemental Material for "Astrocytes: Arnol'd Tongues Generalization in Dynamical Systems' Parameter Plane"

Gonzalo Marcelo Ramírez-Ávila,[1,2,3] S. Leo Kingston,[4] Marek Balcerzak,[2] Jérôme Daquin,[5] Timoteo Carletti,[1] and Tomasz Kapitaniak[2]

[1]*Namur Institute for Complex Systems (naXys),*

*Université de Namur, Rue de Bruxelles 61, B-5000 Namur, Belgium*

[2]*Division of Dynamics, Lodz University of Technology, 90-924 Lodz, Poland*

[3]*Instituto de Investigaciones Físicas, Universidad Mayor de San Andrés, La Paz, Bolivia*

[4]*Centre for Nonlinear and Complex Networks, SRM Institute of Science and Technology, Ramapuram, Chennai 600089, Tamil Nadu, India*

[5]*Univ Toulon, Aix Marseille Univ, CNRS, CPT, Toulon, France*

## S1. COMPLEMENTARY MATERIAL FOR ASTROCYTE 1

Here we outline the origins of the study, which initially focused solely on describing quasiperiodicity but has since evolved to examine the new structures discovered in the parameter plane (PP) analysis of the Zeeman laser dynamic system, including the features of the dynamic properties of astrocyte 1 (time series, attractors, Poincaré maps, computation of Lyapunov exponents (LEs) and other dynamical indicators, and also the self-similar structures emanating from it).

### A. Origin of the work

The origin of this work is based on the PP presented in Fig. 1 of [55], where the authors identified the quasiperiodic breakdown. Figure S1(a) illustrates the same PP, utilizing the number of isospikes to identify the regions characterizing periodic behavior, whereas Fig. S1(b) employs the spectrum of Lyapunov exponents (LEs) to differentiate among the quasiperiodic, chaotic, and hyperchaotic regimes, focusing on the two largest Lyapunov exponents (LLEs). Note that in Fig. S1(b), the pink region is assumed to be hyperchaotic, since it is characterized by two positive LEs, whereas the green olive region is featured by the null values of the two LLEs (in numerical terms, $\lambda_1 \approx 0$ and $\lambda_2 \approx 0$).

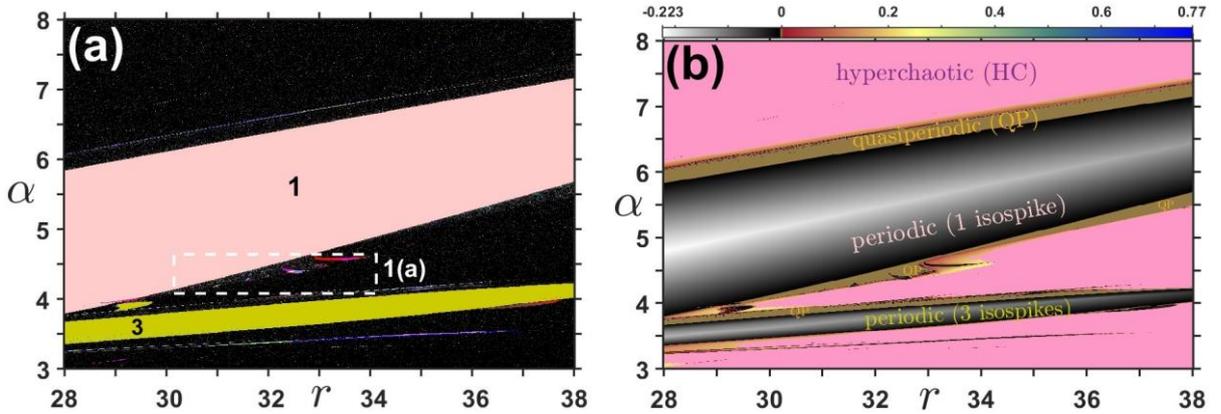

FIG. S1: Parameter plane $\alpha$ vs $r$ with $\sigma = 6.0$ shown in [55] in terms of (a) the number of isospikes of the variable $E_y$, where it is possible to distinguish two large regions of periodic behavior with one and three isospikes. and (b) the two LLEs, where the color bar indicates the value of the LLE ($\lambda_1$) for the chaotic region (colored part) and the second LLE ($\lambda_2$) for the periodic region (black and white part). The green olive part indicates a quasiperiodic (QP) regime (pink region) with $\lambda_1 \approx 0$ and $\lambda_2 \approx 0$. and the hyperchaotic (HC) regime with $\lambda_1 > 0$ and $\lambda_2 > 0$.

### B. Time series and return maps

We show some examples of time series (Figs. S2(a), (c), and (e) with 16, 17, and 18 isospikes per period, respectively) and their return maps (Figs. S2(b), (d), and (f)) corresponding to points located in astrocyte 1 soma's regions, which belong to the first layer in the quint-point cascade.

### C. Attractors in phase portraits

Different 2D and 3D-phase portraits, with different combinations of dynamical variables, are shown to distinguish the diverse types of dynamical behavior. Thus, in Fig. S3 are represented the phase portraits of $E_x$ and $E_y$ (first column) accompanied by their return maps (second column) for the cases in which the system exhibits quasiperiodic, chaotic, and hyperchaotic motion. Whereas, in Fig. S4 are represented the 3D-phase portraits for typical values of each kind of dynamical behavior shown in Table S1. By the computation of the two LLEs ($\lambda_1$ and $\lambda_2$) [60], we consider four kinds of behavior in Fig. S4: periodic (P), quasiperiodic (QP), chaotic (C), and hyperchaotic (HC), according to Table S1 when $\sigma = 6.0$ and considering the situation of a regular oscillation with 15 isospikes. In Fig. S4 we represent the 3D-phase portraits of the linear polarization components of the electric field ($E_k$), the transition coherence ($P_k$), and the atomic inversion ($D_k$) being $k = x,y$: in the left column ($E_x, P_x, D_x$) and in the right column ($E_y, P_y, D_y$) for the 5 cases stated in Table S1. Note the shapes of the attractors in the 3D-phase portraits of Fig. S4, which are concordant with the expected typical dynamic behavior.

TABLE S1: Parameter values, the two LLEs, and the corresponding dynamical regimes when $\sigma = 6.0$.

| $r$ | $\alpha$ | $\lambda_1$ | $\lambda_2$ | behavior |
|---|---|---|---|---|
| 32.4809 | 4.404 | $1.4 \times 10^{-5} \approx 0$ | -0.07522 | P1 |
| 32.302 | 4.36453 | 0.015044 | -0.003 | C1 |
| 32.825 | 4.3556 | 0.249006 | 0.014977 | HC1 |
| 32.2731 | 4.404 | $-1.6 \times 10^{-5} \approx 0$ | $-8 \times 10^{-6} \approx 0$ | QP1 |
| 32.422 | 4.42528 | $9 \times 10^{-6} \approx 0$ | $3 \times 10^{-6} \approx 0$ | QP2 |

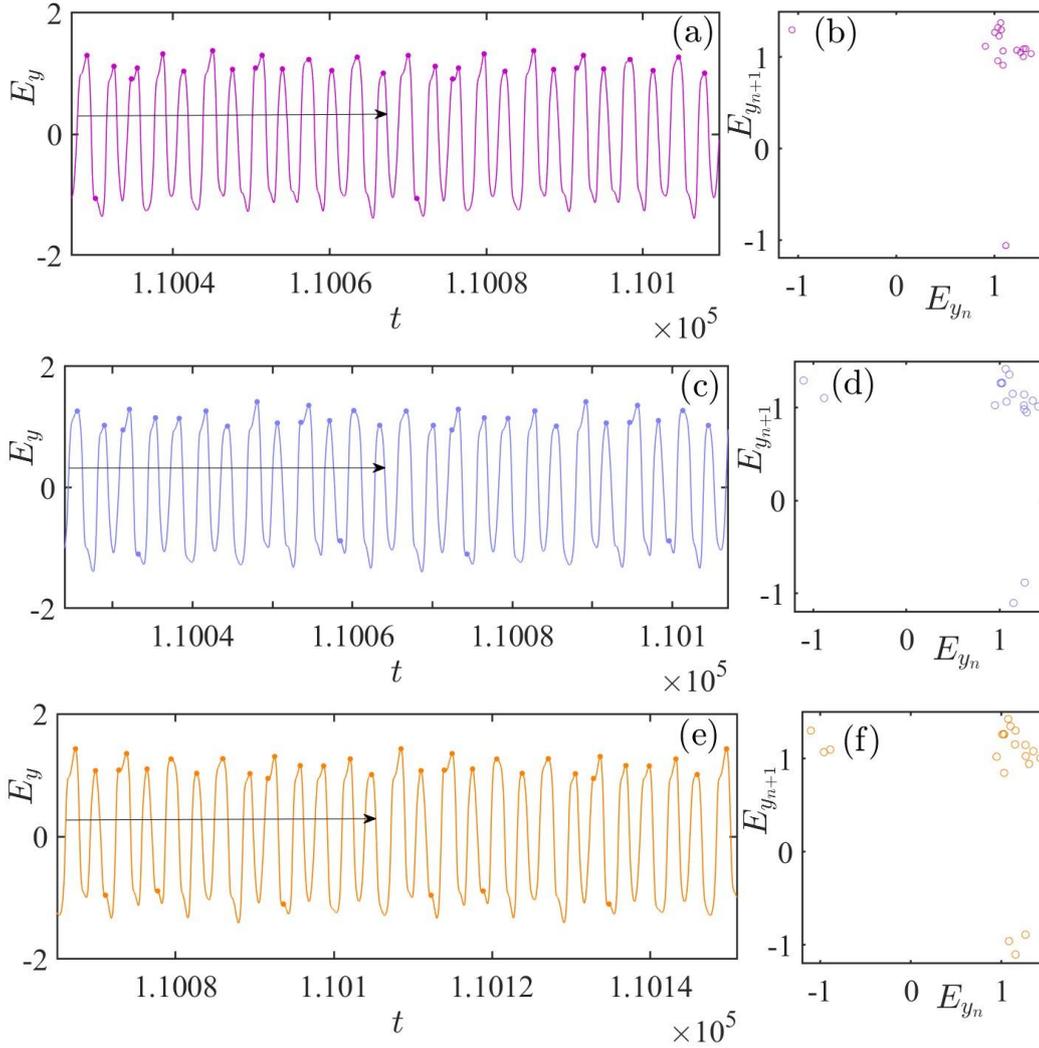

FIG. S2: Time series of the dynamical variable $E_y$ and return maps in the $E_{yn+1}$ versus $E_{yn}$ plane at three representative points in the astrocyte's 1 region exhibiting distinct periodic behavior: (a) and (b) $r$ = 32.4 16 isospikes per period (period-16); (c) and (d) $r$ = 32.4542, 17 isospikes per period (period-17); and (e) and (f) $r$ = 32.502, 18 isospikes per period (period-18), respectively. The return map for each temporal evolution confirms the existence of various periodic dynamics in the Zeeman laser model. In all cases $\alpha$ = 4.3745 and $\sigma$ = 6.0.

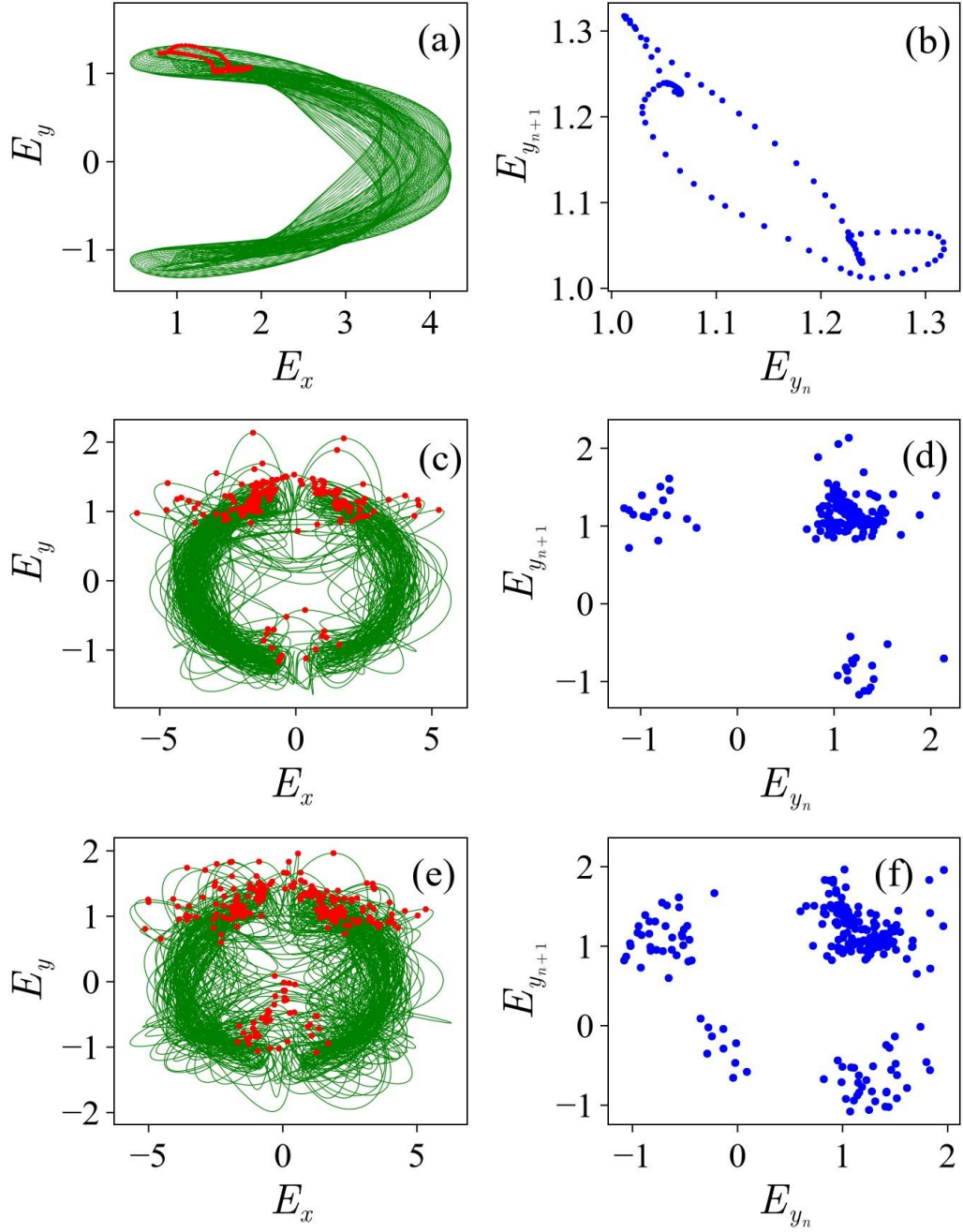

FIG. S3: The phase portraits and return maps illustrate different complex dynamics of the Zeeman laser model: (a) and (b) show quasiperiodic behavior for $\alpha$ = 4.42 and $r$ = 32.25; (c) and (d) depict chaotic attractors for $\alpha$ = 4.39 and $r$ = 32.84; while (e) and (f) represent hyperchaotic dynamics for $\alpha$ = 4.0 and $r$ = 32.0.

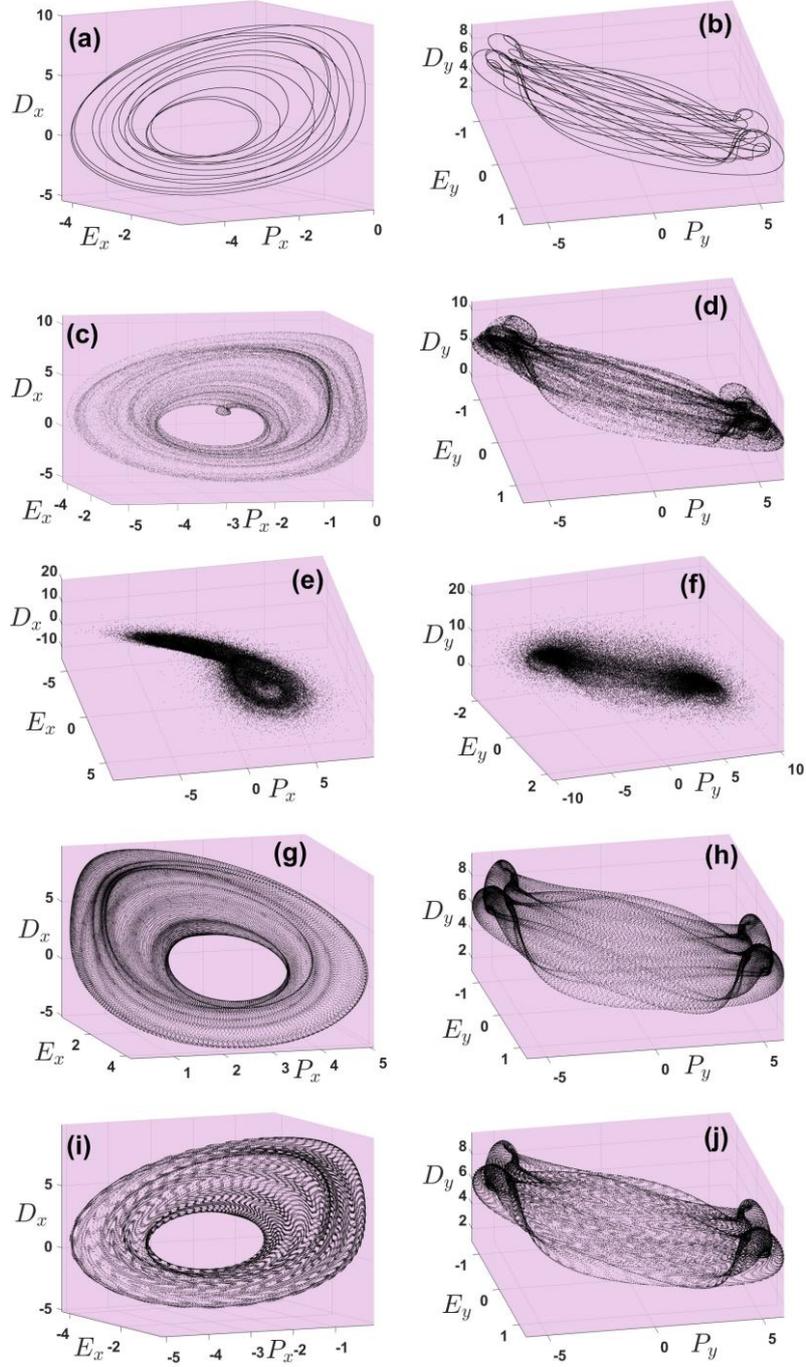

FIG. S4: Phase portraits of the linear polarization components of the electric field ($E_k$), the transition coherence ($P_k$), and the atomic inversion ($D_k$), where $k = x, y$ are the components of the quantities in each direction. First column: ($E_x, P_x, D_x$); second column: ($E_y, P_y, D_y$). (a) and (b) ($\alpha, r$) = (32.4809, 4.404), periodic. (c) and (d) ($\alpha, r$) = (32.302, 4.36453), chaotic. (e) and (f) ($\alpha, r$) = (32.825, 4.3556), hyperchaotic. (g) and (h) ($\alpha, r$) = (32.2731, 4.404) and (g) and (h) ($\alpha, r$) = (32.422, 4.42528) are quasiperiodic.

### D. Evolution of the dynamics indicators

Employing diverse methods to compute the LLEs, such as those proposed by Benettin et al. [65] and Maffione et al. [63], we can appreciate similar results of the LLEs, which permit us to see the differences in the dynamic regimes mentioned in Table S1. Thus, in Fig. S5(a)-(c), the evolution of the two LLEs ($\lambda_1$ and $\lambda_2$) and the mean MEGNO indicator ($\bar{Y}$) are represented, respectively, all computed using MEGNO, and in (d), the evolution of the maximum Lyapunov characteristic exponent (mLCE) computed from the variational equations [62]. Basically, (a) and (d) describe the same quantity, and it is observed that for the case of periodic and quasiperiodic behavior, $\lambda_1$ and mLCE tend to zero and to a positive value when the dynamical behavior is chaotic or hyperchaotic. For $\lambda_2$, the dynamical evolution behavior is similar to that of $\lambda_1$, but with small oscillations around zero for the quasiperiodic cases and a tendency to a finite negative value for the periodic case as it is shown in the inset of Fig. S5(b). The indicator $\bar{Y}$ increases (decreases) as a power law for the chaotic and hyperchaotic (periodic and quasiperiodic) cases as shown in Fig. S5(c).

### E. Emergent self-similar structures

The self-similar structures arising from astrocyte 1 are illustrated in Figs. 3(c)-(f) of the main text. In Fig. S6, we present the PPs using the LLEs for Figs. 3(e)-(f). This complement reinforces our finding that the structures originate from the astrocyte. Thus, the branch and a portion of the soma without any vertices exist in a quasiperiodic scenario; on the contrary, the vertices of the soma lie in a chaotic landscape. In Fig. S6(a), there are two concavities with at least three vertices corresponding to the sequences of quint points observed in Fig. 3(e). Besides these concavities, there are two quasiperiodic regions housing a bunch of Arnold tongues or a new generation of astrocytes-characterizing the regular behavior of the dynamical system. In Fig. S6(b), additional self-similar structures emerge from astrocyte 1, and it is interesting to see a kind of crescent, similar to that shown in Fig. 1(a), and a bundle of Arnol'd tongue-like structures, which are actually another generation of astrocytes. Certainly, the process might continue, extending the offspring featured by smaller sizes, a lesser number of vertices, and periodic behavior with an increasing number of isospikes per period.

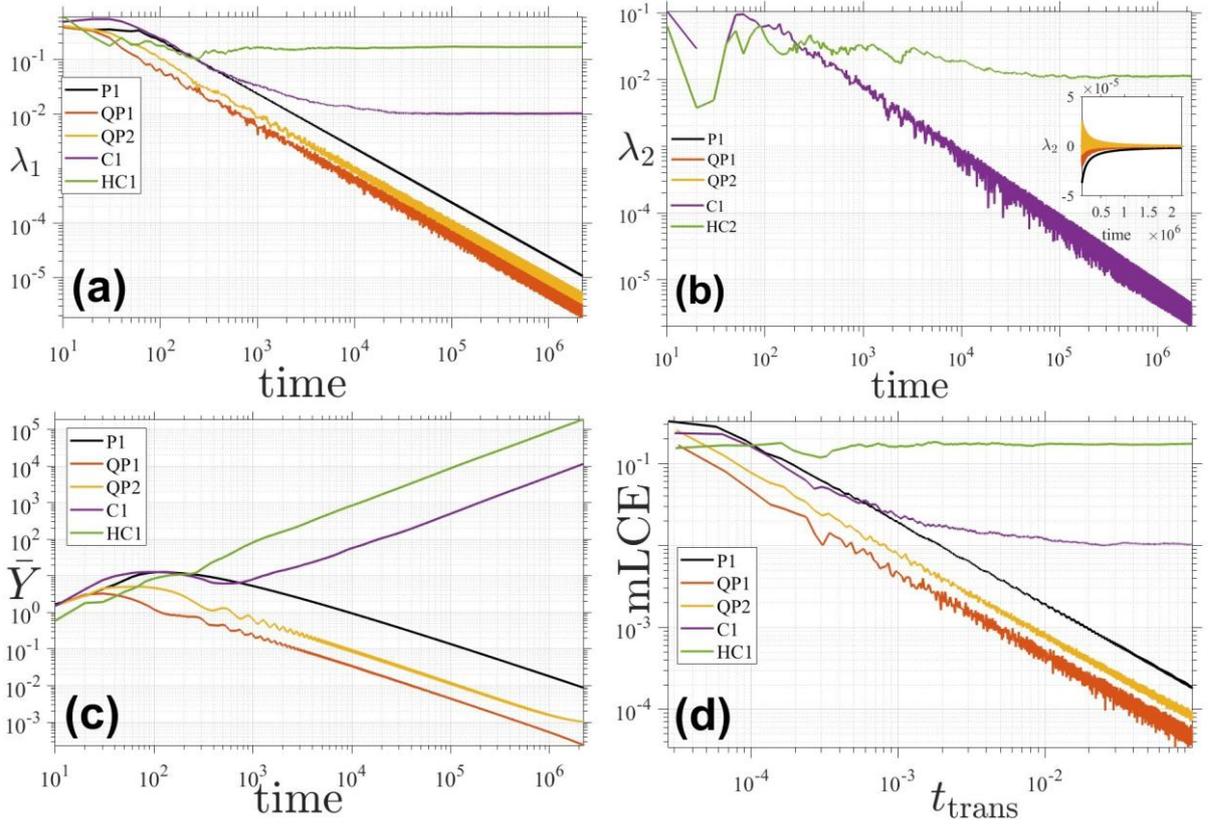

FIG. S5: Evolution of the LLEs using the parameter values indicated in Table S1, $(\alpha,r)$ = (32.4809,4.404), P1 (black); $(\alpha,r)$ = (32.302,4.36453), C1 (violet); $(\alpha,r)$ = (32.825,4.3556), HC1 (green); $(\alpha,r)$ = (32.2731,4.404), QP1 (orange), and $(\alpha,r)$ = (32.422,4.42528), QP2. Computation using MEGNO (a)–(c): (a) First LLE, $\lambda_1$. (b) Second LLE, $\lambda_2$. (c) Mean MEGNO indicator. (d) Evolution of the mLCE derived from the variational equations in terms of the transient.

## S2. STRUCTURES BESIDES ASTROCYTE 1

As illustrated in Fig. 3(a) and 3(b), beside astrocyte 1, there are kite-shaped and horn-shaped Arnol'd tongue structures. These structures exhibit morphology (branch, soma, and vertices) and properties similar to those of astrocytes. Specifically, the branch and the portion of the soma without vertices are in a quasiperiodic environment, while the parts of the soma that contain vertices are immersed in what can be described as a "chaotic ocean." Within the soma, the dynamic behavior of the system is related to periodic behavior characterized by a high number of isospikes per period and the possibility of observing, in a concave region, a sequence of quint-point duplication, similar to a period-doubling route to chaos. In other words, the Arnol'd tongues constitute a particular case of astrocytes but with a reduced number of vertices. The latter observation allows us to classify astrocytes based on their complexity, which includes factors such as branch length, soma size, vertex count, and their generational origin. Figure S7(a) presents an enlarged view of the framed region from Fig. 3(a), where we identify quint points and sequences among them, particularly in the highlighted area, which is further enlarged in (b). The distribution of regions with the same number of isospikes exhibits a certain symmetry from a dynamical perspective, related to the existence of both clockwise and anti-clockwise quint points. These are depicted as circles with arrows indicating the orientation in which the number of isospikes increases. An ideal distribution of periodic regions exhibiting the chirality property characterized by the sequence of the $2^n$ quint points through the $n$ layers is presented in Fig. S7(c). In our case, as shown in Fig. S7(d), there is a symmetry breaking that leads to local chiral behavior in the central and outer edge parts but not to global chirality, as seen in the preceding layers. Figure S7(e) shows the symmetry breaking that leads to global chirality loss and gives rise to symmetric and asymmetric domains.

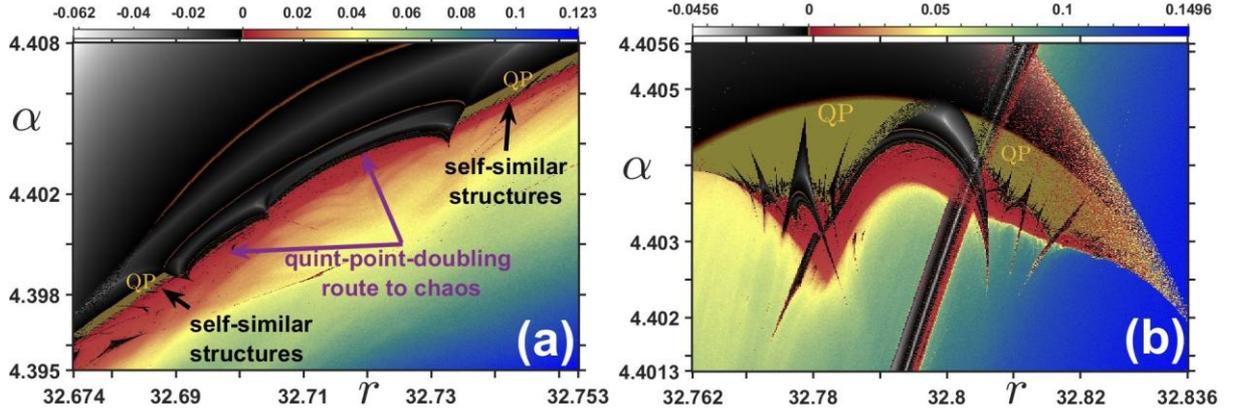

FIG. S6: Parameter planes in terms of the LLEs for the astrocyte 1's emanating structures shown in Figs. 3(e)–(f). (a) Two regions of quasiperidicity (QP) appear besides the concavity. (b) Emerging self-similar astrocytes generating an additional QP region. Note the presence of a crescent similar to that appearing in Fig. 1(a).

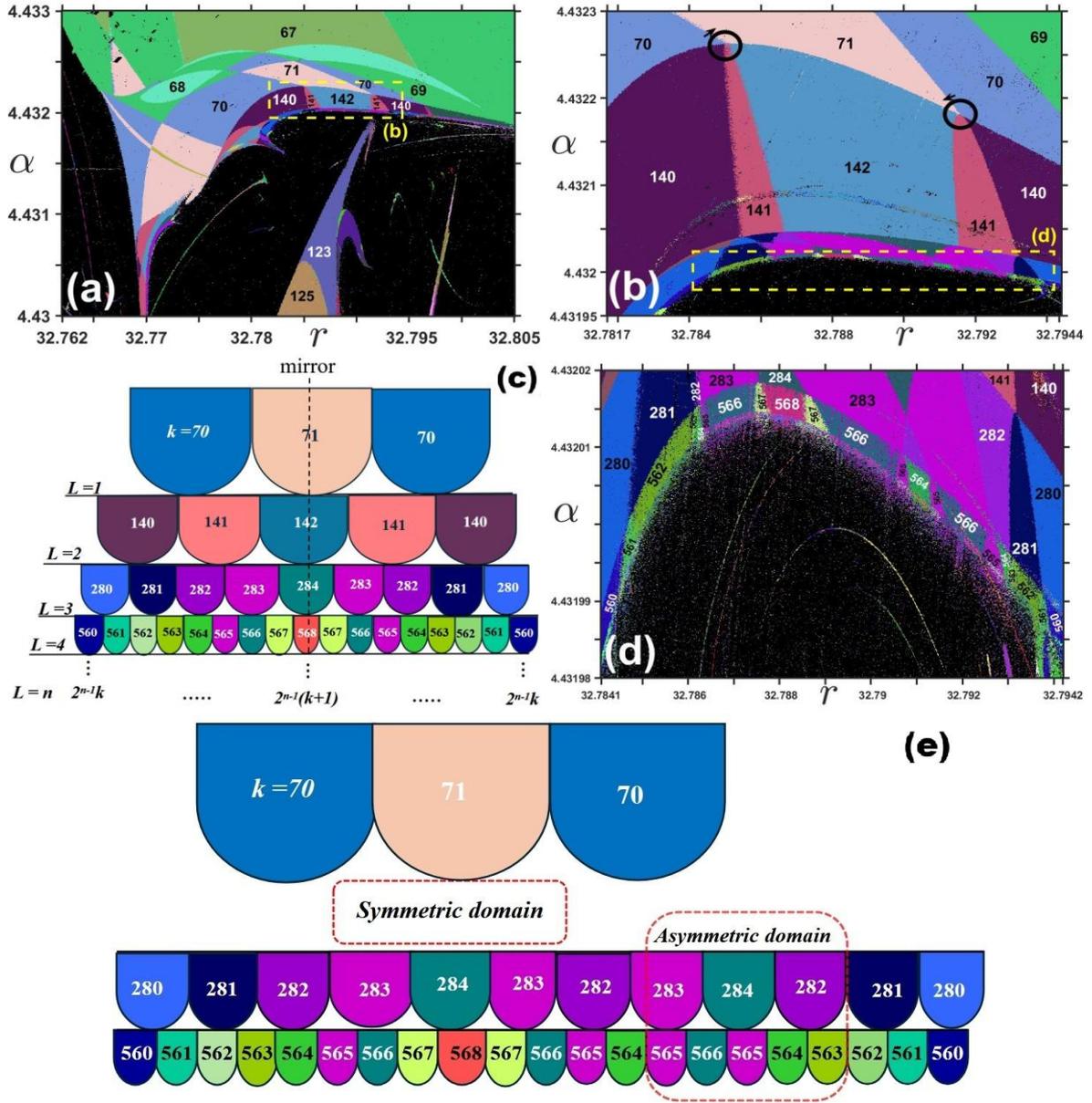

FIG. S7: Parameter planes of the variable $E_y$ in terms of the number of isospikes corresponding to the sequence of quint points observed in the Arnol'd tongue besides astrocyte 1. (a) Concavity showing a sequence of quint points. (b) Magnification of the framed region in (a), where two quint points, one clockwise and the other anti-clockwise, give rise to a chiral property of the distribution of periodic regions characterized by their number of isospikes. (c) An ideal symmetric representation of the layers and chiral distribution of periodicities. (d) Magnification of the framed region shown in (b), where it is possible to distinguish a symmetry breaking in the distribution of periodicities, with the consequent loss of global chirality. (e) Distribution of periodicities when symmetric and asymmetric domains appear as a result of global loss of chirality.

## S3. COMPLEMENTARY MATERIAL FOR ASTROCYTE 2

We consider the framed regions of Fig. 1(e) both exhibiting tri (cyan), quart (magenta) and quint (yellow) point sequences, and the corresponding layers in their concave parts as shown in Fig. S8(a)–(b). In the astrocyte's inner part, we note from Fig. 1(e) that the basic periodicity is 17 and other periodicities such as 19-23 pervade most of the soma's area and form the typical harlequin dress-like pattern. In Fig. S8(c) is the magnification of the framed region of Fig. S8(a) showing the quint point sequence in the route to chaos. On the other hand, the periodic arcs also appear as for astrocyte 1. Finally, in Fig. S8(d), is the zoom of the framed region of Fig. S8(c) where an exotic structure with a similar shape that a manta-ray lies on the chaotic region and exhibits a rich dynamic behavior, including the sequence of quint points, several vertices and even the "eye" of chaos.

## S4. ASTROCYTES IN OTHER DYNAMICAL SYSTEMS

Here, we show some examples of astrocyte-like structures seen in discrete and continuous systems, and this reinforces our claim that astrocytes and their features constitute a general and unified approach for describing the regions commonly known as Arnol'd tongues.

### A. Modified Hénon map

We consider a modified version of the Hénon map [68] given by:
$$x_{n+1} = a - bx_n^2 - cy_n^2 \quad \text{(S1)}$$
$$y_{n+1} = x_n.$$

The PP $a \times b$, with $c = 1$, in terms of periodicities is shown in Fig. S9(a), where we note the presence of the astrocytes with the main features stated in the Letter, i.e., a branch emerging from a region describing regular dynamical behavior (white region), more specifically a fixed point, and a soma with several vertices; the branch and the part of the soma without vertices are submerged in a region describing quasiperiodicity and the vertices of the astrocyte lying in the chaotic region, the pink region stands for divergent behavior and that in olive green represents the region characterizing quasiperiodicity. We note the presence of a structure similar to the crescent observed in the Zeeman's laser model PP; it is interesting to see that this structure establishes a kind of division in the PP in a part that increases the astrocytes' periodicities downwards and the other upwards. Figure S9(b), is the PP based on the LLE and it was also the base to distinguish the quasiperiodic region in Fig. S9(a).

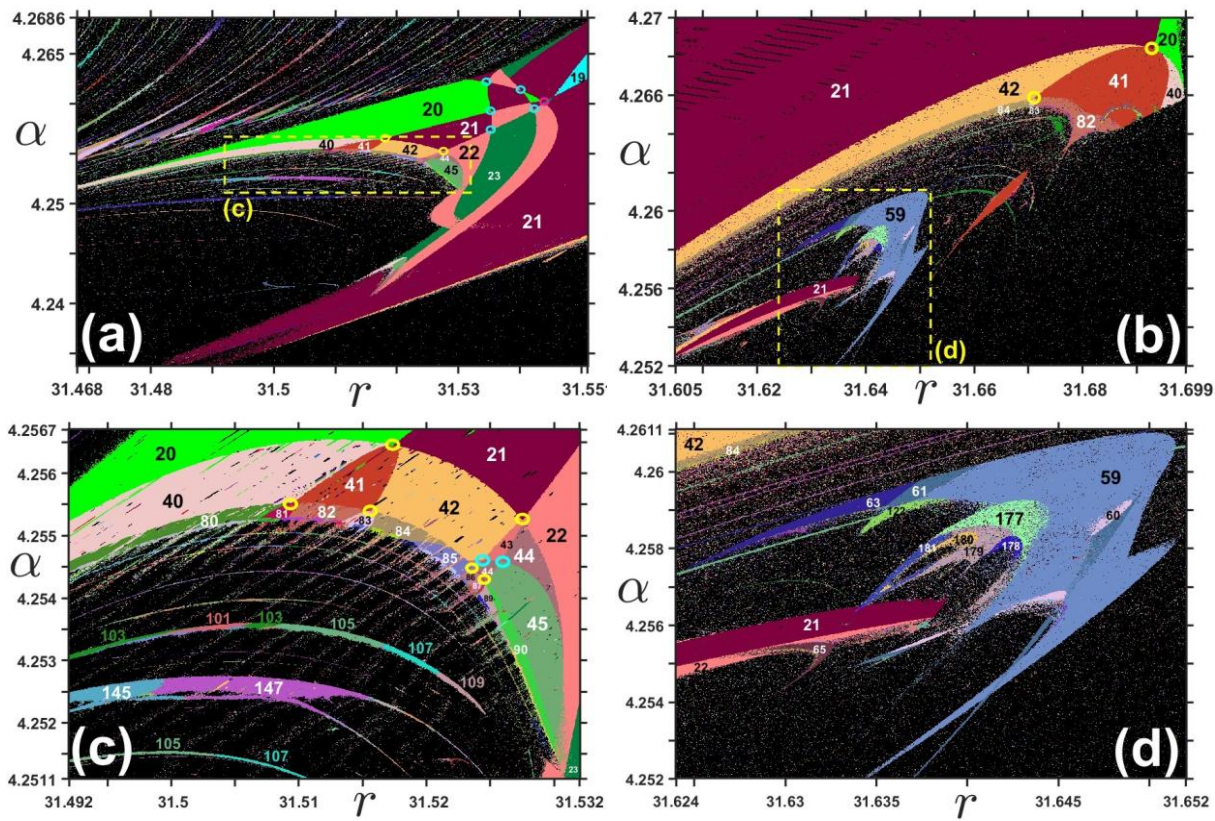

FIG. S8: Parameter planes of the variable $E_y$ in terms of the number of isospikes in the inner and outer parts of astrocyte 2. Enlarged view of the framed regions of Fig. 1(e), wherein there is a sequence of quint points and three visible layers with (a) $L_1$: 20 – 21 –22, $L_2$: 40 – ⋯ – 44 and $L_3$: 80 – ⋯ – 88, and (b) $L_1$: 21 – 20, $L_2$: 42 – 41 – 40, and $L_3$: 84 – 83 – 82, where it is noted that the sequence in $L_3$ is truncated; two quint points are identified with yellow circles. A manta-ray like structure is selected to be zoomed in (d)(c) Magnification of the framed region in (a) showing the four quint points (yellow circles), two tri points (cyan circles) and the third layer $L_3$ with more detail. (d) Manta-ray like structure exhibiting several vertices and the so-called "eye" of chaos.

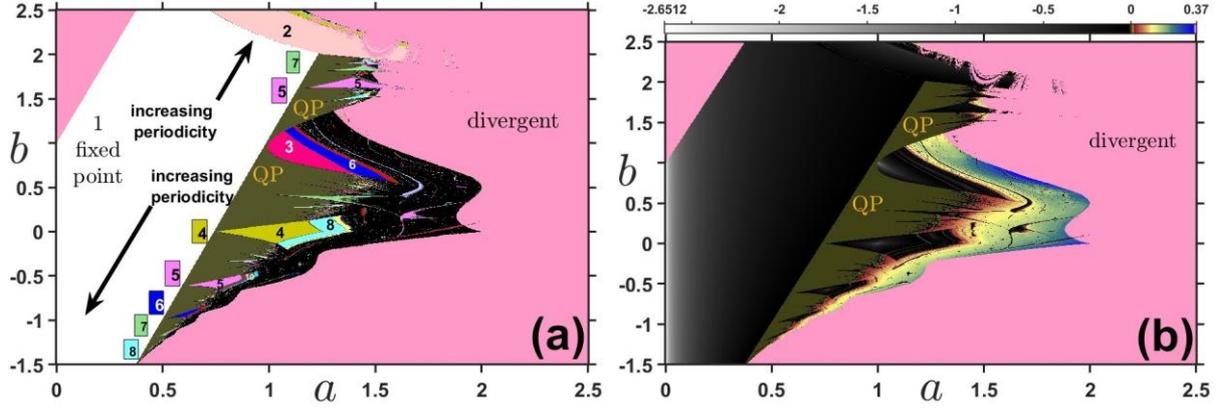

FIG. S9: (a) Parameter plane of the variable *x* in terms of the periodicities and (b) Largest Lyapunov Exponent of the map

### B. Vertical-cavity surface emitting laser (VCSEL)

Another system in which is possible to observe astrocytes is the VCSEL described dynamically in detail in terms of PPs [31], whose equations are given by Eqs. (S2).

$$\begin{aligned}
\frac{dE_x}{dt} =& \kappa(1+i\alpha)(NE_x + inE_y - E_x) - i(\gamma_p + \Delta\omega)E_x \\
& - \gamma_a E_x \quad, \\
\frac{dE_y}{dt} =& \kappa(1+i\alpha)(NE_y - inE_x - E_y) + i(\gamma_p - \Delta\omega)E_y \\
& + \gamma_a E_y + \kappa_{\text{inj}} E_{\text{inj}} \quad, \\
\frac{dN}{dt} =& -\gamma_e \left[ N\left(1 + |E_x|^2 + |E_y|^2\right)\right] + \gamma_e \mu \\
& - i\gamma_e n \left( E_y E_x^* - E_x E_y^* \right) \quad, \\
\frac{dn}{dt} =& -\gamma_s n - \gamma_e n \left(|E_x|^2 + |E_y|^2\right) \\
& - i\gamma_e N \left( E_y E_x^* - E_x E_y^* \right) \quad.
\end{aligned} \quad (S2)$$

In Fig. S10(a) it is shown a bunch of astrocytes emerging from the vast region characterizing periodic dynamical behavior with three isospikes per period with a sequence for the main astrocytes: 21 – 33 – 45 – 57 – ... is shown, and it is possible to generalize for this case to the sequence for the main astrocytes: $n_m = (n_{\text{source}} + 4 \times (n_1 + m - 1) \times n_{\text{source}}$, where $n_1$ is the number of isospikes per period or periodicity of the first astrocyte hierarchically speaking; $n_{source}$ is the periodicity of the structure from which the astrocytes emerge, and $n_m$, the number of isospikes of the astrocyte with hierarchy $m$, being in our case: $n_1 = 7$, $n_{source} = 3$. In Fig. S10(b), we have a zoomed view of the main astrocyte showing several vertices and multiple periodicities in its soma.

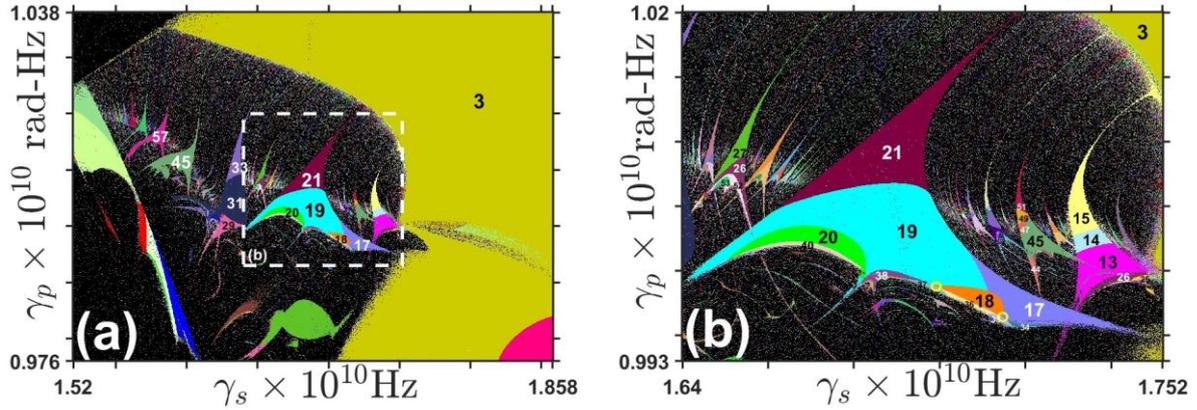

FIG. S10: (a) Parameter plane $\gamma_s \times \gamma_p$ of the variable $n$ in terms of the number of isospikes, where it is noted the bunch of emergent astrocytes from a region in which the dynamical behavior is characterized by oscillations with three isospikes and exhibits a sequence of number of isospikes: 21 – 33 – 45 – 57 – ... for the main astrocytes. (b) Enlarged view of the main astrocyte and, in the soma, the typical harlequin dress pattern, indicating the possibility of quint-point doubling cascades, i.e., a bifurcation within the PP.